\documentclass[aps,prd,amsmath,amssymb,10pt,letterpaper,balancelastpage,showpacs,superscriptaddress,nofootinbib,floatfix,notitlepage,longbibliography,twocolumn
]{revtex4-2}
\usepackage[utf8]{inputenc}
\usepackage{bm}

\newcommand{\pbh}{\rm MACHO}
\usepackage{comment}
\usepackage{slashed}
\usepackage{latexsym}
\usepackage{graphics}
\usepackage{bm}
\usepackage{color}
\usepackage{amsmath}
\usepackage{amssymb}
\usepackage{mathrsfs}
\usepackage{graphicx}
\usepackage{slashed}
\usepackage{siunitx}
\usepackage{braket}
\usepackage{calligra}
\usepackage{caption}
\usepackage{subcaption}

\usepackage{caption}
\usepackage{subcaption}

\begin{document}
\title{Constraints on Dark Matter from Dynamical Heating of Stars in Ultrafaint Dwarfs. Part 2: Substructure and the Primordial Power Spectrum}

\date{\today}

\author{Peter W.~Graham} 
\email{pwgraham@stanford.edu}
\affiliation{Stanford Institute for Theoretical Physics, Department of Physics, Stanford University, Stanford, CA 94305, USA}
\affiliation{Kavli Institute for Particle Astrophysics \& Cosmology, Department of Physics, Stanford University, Stanford, CA 94305, USA}

\author{Harikrishnan Ramani} 
\email{hramani@udel.edu}
\affiliation{Stanford Institute for Theoretical Physics, Department of Physics, Stanford University, Stanford, CA 94305, USA}
\affiliation{Department of Physics and Astronomy,
University of Delaware, Newark, DE 19716, USA}

\begin{abstract}
There is a large and growing interest in observations of small-scale structure in dark matter.
We propose a new way to probe dark matter structures in the $\sim 10 - 10^8 \, M_\odot$ range.
This allows us to constrain the primordial power spectrum over shorter distances scales  than possible with direct observations from the CMB.
For $k$ in the range $\sim 10 - 1000 \, {\rm Mpc}^{-1}$ our constraints on the power spectrum are  orders of magnitude stronger than previous bounds.
We also set some of the strongest constraints on  dark matter isocurvature perturbations.
Our method relies on the heating effect such dark matter substructures would have on the distribution of stars in an ultra-faint dwarf galaxy.
Many models of inflation produce enhanced power at these short distance scales and can thus be constrained by our observation.
Further, many dark matter models such as axion dark matter, self-interacting dark matter and dissipative dark matter, produce dense structures which could be constrained this way.
 
\end{abstract}

\maketitle

\section{Introduction}

The nature of dark matter (DM) is one of the biggest outstanding mysteries in physics.  We have a wide variety of probes of dark matter including direct and indirect detection and collider experiments.
The study of dark matter structure in the universe is another important probe. Dark matter over-densities collapse in the early universe leading to non-trivial structure over several mass and length scales including clusters, galaxies and dwarf galaxies. The matter power spectrum obtained from correlations in the observed structure agrees very well with the $\Lambda$CDM paradigm down to the $\approx \textrm{Mpc}$ scale.  However there has been a large and growing interest recently in studying shorter scales, as a great deal can be learned both about the nature of dark matter and about the formation of the universe.
While a scale invariant primordial spectrum of density perturbations should lead to formation of structures at all length scales, there are many possible cosmological scenarios which produce even more structure at these shorter length scales.
For example many models of inflation produce enhancements in the primordial power spectrum at short scales, causing additional, denser substructures to form (see e.g.~\cite{Leach:2001zf, Pajer:2013fsa, Hertzberg:2017dkh, Mishra:2019pzq, Inomata:2021tpx, Hooper:2023nnl}).  
Additionally a period of early matter dominance can enhance small-scale power (see e.g.~\cite{Erickcek:2011us, Fan:2014zua, Erickcek:2015jza, Dror:2017gjq}).
Further, even with a scale-invariant initial power spectrum, nontrivial dynamics in the dark sector can cause substructures to form.  For example, axion dark matter, dark photon dark matter, self-interacting dark matter (SIDM), atomic dark matter, dissipative dark matter, and mirror sectors can form dense substructures (see e.g.~\cite{Kolb:1993zz, Visinelli:2017ooc, Eggemeier:2019jsu, Arvanitaki:2019rax, Graham:2015rva, Gorghetto:2022sue, Gilman:2021sdr, Slone:2021nqd, Zeng:2023fnj, Gad-Nasr:2023gvf, Mace:2024uze, Gemmell:2023trd, Roy:2023zar, Chang:2018bgx, Curtin:2019ngc, Dvali:2019ewm}).  Thus, studying  dark matter structures on these short scales can teach us much about the origin of the universe and the nature of dark matter.

The study of dark matter structures becomes increasingly difficult at smaller length/mass scales below dwarf galaxy size, owing to the inability of these dark matter halos to produce significant numbers of stars.  Techniques to study these structures have instead involved gravitational lensing~\cite{Zumalacarregui:2017qqd,dhawan2023type, DeRocco:2023hij} of light, photometric lensing~\cite{VanTilburg:2018ykj} and stellar streams \cite{Banik:2019cza, Ando:2022tpj}.
These promising approaches will hopefully reveal much about the small-scale structure of dark matter in the future.
In this paper we describe a new, complementary approach to studying dark matter structures in the mass range $\sim 10 - 10^8 \, M_\odot$, or primordial density perturbations in the range of $k$ from $\sim 10 - 1000 \, {\rm Mpc}^{-1}$.

These smaller structures, which we refer to as clumps in this work, collapse before galaxies and subsequently fall into the galaxies as they form. Thus, these clumps make up intra-galactic substructure and would survive until today barring destruction due to tidal forces present in the host. 
In this work, we describe an important probe of such sub-structure down to the $\approx 10 M_\odot$ mass scale, specifically due to their presence inside some of the faintest substructure identified yet: ultra-faint dwarf galaxies (UFDs).

Part one of this two-part paper set stringent limits on MACHOs/PBHs making up some or all of the UFD dark matter~\cite{Graham:2023unf}. These limits were obtained via considering the gravitational scattering of MACHOs off stars (for similar works see e.g.~\cite{Koulen:2024emg, Brandt:2016aco, Lu:2020bmd, Koushiappas:2017chw, zhu2018primordial, Stegmann:2019wyz, zoutendijk2020muse, Wadekar:2022ymq}).  Such scattering will, on average, transfer heat from the MACHO's kinetic energy to the stellar kinetic energy which would result in the expansion of the stellar scale radius.  This can be compared to the observed stellar scale radius to set constraints.  That paper was essentially in the `point-mass' limit, taking the MACHOs/PBHs to be arbitrarily small. In this sequel we extend the analysis to diffuse dark matter clumps of arbitrary density. 
Importantly, we still expect a similar heat transfer rate from the dark matter clumps to the stars even if the dark matter clumps are significantly larger than the MACHOs, and in fact even if the dark matter clump size approaches the size of the stellar scale radius ($\sim 30 \, {\rm pc}$).  This is because the minimum impact parameter for gravitational clump-star scattering enters only via a log.  Thus even though moving from point-like MACHOs to diffuse clumps will cause an increase by many orders of magnitude in the minimal impact parameter, this only enters the log in the scattering and so has very little effect on the heat transfer rate.
Thus this effect should be a powerful probe of even relatively diffuse DM substructures.  However for low enough densities, such clumps will not survive for long in the environment of the UFD, thus cutting off the heating effect at some minimal density for the clumps.
In this work we analyze the survival prospects of these diffuse objects in the UFD environment. For the clumps that survive, we estimate the effects of the finite size of the clumps on the heat transfer rate thus deriving limits on the existence of clumps in the UFD.   

The rest of this paper is organized as follows. Section~\ref{clumpheat} describes our calculation of the gravitational heat transfer rate from clumps to stars. Section~\ref{clumpsurvival} analyzes the survival probability of clumps against various disruption processes. We present results in Section~\ref{results} and discuss consequences in Section~\ref{sec:discussion}.

\section{Heat Transfer for Clumps}
\label{clumpheat}

In this section we describe our assumptions about the clumps that are inside the UFD and compute the heat transfer rate due to gravitational scattering with stars as well as the smooth DM. 

\subsection{Parametrization of clumps}

First, we consider dark matter to reside in subhalos all of the same mass $M_{\rm clump}$. 
While this assumptions simplifies the limit setting procedure, the limit calculation is straightforward for extended halo mass functions which we leave for future work. 

We consider only NFW clumps and characterize them as 

\begin{align}
\rho_{\rm NFW}(r)=4\rho_{s}\left(\frac{r}{r_s}\right)^{-1}
   \left(\frac{r}{r_s}+1\right)^{-2}
\end{align}
Here, $r_s$ is the scale radius, $\rho_s$ is the density at the scale radius. Owing to the fact that the NFW cores are more robust to tidal stripping, we follow~\cite{VanTilburg:2018ykj} and truncate the NFW profile at the scale radius $r_s$ with masses 
\begin{equation}
    M_{\rm clump}=8\pi r_s^3 \rho_s (-1 + \log 4)
\label{denseqn}
\end{equation}
These cores are assumed to make up a fraction $f_{\rm clump}$ of the total dark matter mass of the UFD. The rest of the dark matter is assumed to be smooth for simplicity. Thus, the three parameters $\{M_{\rm clump},\rho_s,f_{\rm clump}\}$ fully describe the clumps we consider. 
For later use we define the mass enclosed inside a radius $b$ as
\begin{align}
    M_{\rm enc}(b)=\int_0^b 4\pi r^2 \rho_{\rm NFW}(r)
\end{align}
since the clumps are truncated at $r_s$, we assume that $M_{\rm enc}(b>r_s)=M_{\rm clump}$.

Finally, we assume the clumps themselves are distributed following a Dehnen profile in the UFD.  So the averaged mass density of all the clumps in the UFD is: 
\begin{align}
    \rho_{\rm clumps}(r,t)=f_{\rm clump}\frac{3 M_{\rm UFD}}{4\pi R_{\rm clumps}^3(t)} \left(1+\frac{r}{R_{\rm clumps}(t)}\right)^{-4}
    \label{eqn:rhomacho}
\end{align}
with the scale radius $R_{\rm clumps}$ being time dependent due to heat transfer between clumps and the smooth DM component causing the clumps to migrate inwards as derived in the next subsection.  The smooth DM itself also follows a Dehnnen profile but with a constant scale radius $R_{\rm UFD}=446.3~\textrm{pc}$. \begin{align}
    \rho_{\rm DM,smooth}(r,t)=\left(1-f_{\rm clump}\right)\frac{3 M_{\rm UFD}}{4\pi R_{\rm UFD}^3}  \left(1+\frac{r}{R_{\rm UFD}}\right)^{-4}
    \label{eqn:rhosmooth}
\end{align} We assume the smooth DM component is the dominant component of DM so technically our results would not apply if $f_{\rm clump}$ is too close to 1. We calculate the time dependent $R_{\rm clumps}(t)$ next.

\subsection{Migration of clumps}

Although our main effect is the heating of stars, in this subsection we consider the heat transfer from clumps to the dominant, diffuse component of the DM.  This is the dominant heat loss mechanism for the clumps.  It causes the clumps to migrate inwards towards the center of the UFD, decreasing the scale radius of the clump distribution over time.  This increases the density of the clumps at the location of the stars, and is thus an important effect to consider for the heating of the stars.

In Paper 1~\cite{Graham:2023unf} we considered the scenario that dark matter had two components: MACHOs (with mass $M_{\pbh}$) and diffuse DM (with mass $m_{\rm DM}$).  The heat transfer rate from MACHOs to diffuse DM  per unit MACHO mass, in the $m_{\pbh} \gg m_{\rm DM}$ limit was given as 
\begin{align}
    H_{\pbh}
    =2\sqrt{2\pi} G^2  \left(1- f_{\pbh}\right)\rho_{\rm DM}M_{\pbh}\nonumber \\\times \frac{ \sigma_{\pbh}^2}{\left(\sigma_{\rm DM}^2+\sigma_{\pbh}^2\right)^\frac{3}{2}}\log \left(\frac{b_{90}^2+b_{\rm max}^2}{b_{90}^2+b_{\rm min}^2 }\right)
    \label{eqn:heateqndirmacho}
\end{align}
Here we define 
\begin{align}
\rho_{\rm DM}\equiv\rho_{\rm DM,smooth}\left(r\ll R_{\rm UFD}\right)\approx \frac{3 M_{\rm UFD}}{4\pi R_{\rm UFD}^3}
\end{align}
Eqn.~\ref{eqn:heateqndirmacho} assumes the MACHOs are effectively point masses (as would be the case for primordial black holes for example).

In this paper we wish to consider what happens when we replace the MACHOs with finite-size dark matter clumps (bound together by gravity).
So we now make modifications to the above expressions in order to capture the finite size effects of the clump.
First $b_{\rm max}$ is set to the current scale radius of the  distribution of clumps in the UFD, $R_{\rm clumps}(t)$, because this is roughly the orbit size of a typical clump. For point particles, $b_{\rm min}$ was set to zero because there are no sampling limitations for smooth DM. For the present case case, we can conservatively restrict the impact parameter to be larger than the scale radius of each clump, such that any DM particle sees the whole mass of the clump $M_{\rm clump}$ i.e. we set  $b_{\rm min}=r_s$. Hence, we have, 
\begin{align}
    H_{\rm clump}
    =2\sqrt{2\pi} G^2  \left(1- f_{\rm clump}\right)\rho_{\rm DM} M_{\rm clump}\nonumber \\
    \times \frac{\sigma_{\rm clump}^2}{\left(\sigma_{\rm DM}^2+\sigma_{\rm clump}^2\right)^\frac{3}{2}}
   \log \left(\frac{b_{90}^2+R_{\rm clumps}(t)^2}{b_{90}^2+r_s^2 }\right)
    \label{eqn:heateqndirclump}
\end{align}

We then integrate this heat transfer rate over the distribution of clumps to find the total rate of energy loss of the clumps:
\begin{align}
     \frac{d\mathcal{E}_{\rm clump}(R_{\rm clumps})}{dt}= \int dr \, 4\pi r^2  \rho_{\rm clumps}(r) \,\nonumber \\ \times H_{\rm clump} \left( \sigma_{\rm clump}(r),\sigma_{\rm DM}(r),\rho_{\rm DM}(r) \right) 
    \label{heatintmacho}
\end{align}
And then we plug in the expression for the energy of the clumps $\mathcal{E}_{\rm clump}$ which is identical to $\mathcal{E}_{\pbh}$ in Ref.~\cite{Graham:2023unf} with $M_{\pbh}$ set by $M_{\rm clump}$. This is because the potential energy and kinetic energy of an individual clump due to its presence in the UFD should be the same as the corresponding MACHO since the clump radius $r_s$ is much smaller than the size of the UFD, $R_{\rm UFD}$. The initial condition is given by $R_{\rm clumps}(t=0)=R_{\rm UFD}$.  And we can then solve for the evolution of the scale radius of the clump distribution in time, $R_{\rm clumps}(t)$.

What we really care about is the enhancement in the number density of clumps seen by the stars (i.e. at the stellar scale radius $R_{0,\star}$) averaged over the age of the UFD, $T_{\rm UFD}$ compared to the average DM density without the effect of migration.  This is given by 
\begin{align}
\eta_{\rm NT}=\frac{1}{T_{\rm UFD}}\int_0^{T_{\rm UFD}} dt \frac{\rho_{\rm clumps}(R_{0,\star}(t),t)}{\rho_{\rm clumps}(R,t=0)}
\label{eqn:enhancement factor}
\end{align}
Here, the subscript NT emphasizes that no tidal effects have been included. 
Note that $R_{\rm clumps}(t)$ starts at $R_{\rm UFD}$ at $t=0$ and stops contracting when it reaches~\cite{Graham:2023unf} $R_{\rm clumps}^{\rm cutoff}\approx f_{\rm clump}^\frac{1}{3} R_{\rm UFD}$ i.e. when the density of clumps matches the ambient smooth DM density. Note that almost the entire range of $10^{-4} < f_{\rm clump} <1$ that we consider in this paper corresponds to $R_{\rm clumps}^{\rm cutoff}>24.73~\textrm{pc}$. Furthermore, since $R_{0,\star}<24.73~\textrm{pc}$ during its evolution, it is true that $R_{0,\star}< R_{\rm clumps}$, i.e.~the stars are always inside the region containing the maximal density of clumps. Since the core density inside $R_{\rm clumps}$ is roughly a constant owing to the assumption of a Dehnen profile with $\gamma=0$, we can make the approximation
\begin{align}
\rho_{\rm clumps}(R_{0,\star}(t),t)\approx f_{\rm clump}\frac{3  M_{\rm UFD}}{4\pi R_{\rm clumps}^3(t)}
\label{eqn:coreapprox}
\end{align}
In other words, we define the enhancement $\eta_{\rm NT}$ at the stellar scale radius $R_{0,\star}$ but it would be the same when evaluated at any radius inside the scale radius of the clumps. 

\begin{figure}
     \centering
         \centering
         \includegraphics[width=0.48\textwidth]{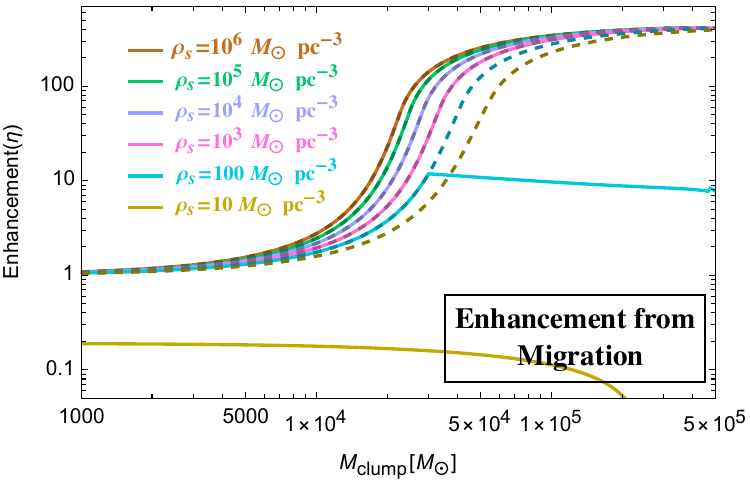}
         \caption{The enhancement factor $\eta(R = 24.7~\textrm{pc})$. This is the enhancement of the clump density at the location of the stars (averaged over the age of the UFD).  The enhancement arises from the heat transfer from clumps to diffuse DM which causes the clumps to migrate inwards towards the center of the UFD over time. The enhancement ignoring clump destruction $\eta_{\rm NT}$  as given in Eqn.~\eqref{eqn:enhancement factor} is plotted in dashed lines. The effective enhancement $\eta_{\rm HHH}$ as given in Eqn.~\eqref{eqn:ehhh} is plotted in solid lines.}
         \label{fig:enhancement}
\end{figure}

The result of solving the differential equation for $R_{\rm clumps}(t)$ and evaluating $\eta_{\rm NT}$ is plotted in Fig.~\ref{fig:enhancement} as a function of the clump mass $M_{\rm clump}$ for different choices of the density of the clump $\rho_s$ for $f_{\rm clump}=10^{-3}$ in dashed lines. The solid lines are the result after clump survival in tidal heating is included and will be discussed in Sec.~\ref{subsec:hhh}.  Note that the $\rho_s$ dependence comes through $r_s$ in Eqn.~\ref{eqn:heateqndirclump}. Since $r_s$ only appears in the $\log$ in Eqn.~\ref{eqn:heateqndirclump}, the enhancement is only weakly dependent on the clump density $\rho_s$. A five order of magnitude drop in the clump density from $10^6 \, M_\odot \, \textrm{pc}^{-3}$ to $10 \, M_\odot \, \textrm{pc}^{-3}$ can be roughly compensated by a factor of 2 increase in the clump mass to achieve the same enhancement.    

Thus we now know what to take for the local density of clumps when we evaluate the heating of the stars by the clumps.

\subsection{Direct heating rate of stars by clumps}

In this subsection we calculate the direct heat transfer rate from clumps to stars.  By direct heat transfer we mean clumps that pass directly through the stellar scale radius (as opposed to tidal heating which we consider in the next subsection).

In Paper 1~\cite{Graham:2023unf} the heat transfer between stars (typical mass $m_\star$) and MACHOs (taken to be point objects) per unit stellar mass, in the $m_{\pbh} \gg m_\star$ limit, was given as 
\begin{align}
    H_\star
    =2\sqrt{2\pi} \rho_{\rm MACHO}(t)  \frac{G^2   
    m_{\pbh} \sigma_{\pbh}^2
    }{\left(\sigma_\star^2+\sigma_{\pbh}^2\right)^\frac{3}{2}} \nonumber \\ \times \log \left(\frac{b_{90}^2+b_{\rm max}^2}{b_{90}^2+b_{\rm min}^2}
   \right) 
    \label{eqn:heateqn}
\end{align}
where $\eta$ is the enhancement in local number density due to migration, $b_{\rm max}=3R_{0,\star}$ and $b_{\rm min}=b_{\rm samp}$ and here $R_{0,\star}$ is the stellar scale radius and $b_{\rm samp}$ is the sampling radius, i.e.~the smallest radius around a star through which at least 3 MACHOs will pass in the age of the UFD.
We now modify this formula to change from MACHOs (point particles) to extended clumps with scale radius $r_s$. 

In the limit $b_{\rm max}>b_{\rm samp}>r_s$, there is no need for any modification since for scattering purposes, the clump is a point particle due to Newton's shell theorem.  We just make the obvious replacements $m_{\pbh} \rightarrow  M_{\rm clump}$, $f_{\pbh} \rightarrow  f_{\rm clump}$ and $\sigma_{\pbh} \rightarrow  \sigma_{\rm clump}$.

When $b_{\rm max}>r_s>b_{\rm samp}$, we can conservatively take $b_{\rm min}=r_s$, requiring the star to pass outside the dark matter clump.  Although there will be some extra heating effect from scatterings where the star passes through the dark matter clump, we ignore this extra heating effect which gives us a conservative bound.

Finally, when $r_s>b_{\rm max}$, we can conservatively choose to use the heating effect of only a smaller part of each clump (the part within a certain radius $b_{\rm opt}$ of its center).  We choose this radius to be small enough that this part of the clump does fit within the stellar distribution so  $b_{\rm opt} < b_{\rm max}$.  This is conservative because we are only using the heating effect of a fraction of each clump, the rest of the clump will add some extra heating but we neglect that to find a conservative bound.  Thus for impact parameters between $b_{\rm opt}$ and $b_{\rm max}$ we can still use our old point mass formula Eqn.~\eqref{eqn:heateqn}.  We thus take the minimum impact parameter to be $b_{\rm min}=b_{\rm opt}$.  Then we also have to replace $M_{\rm clump} \rightarrow M_{\rm enc}(b_{\rm opt})$ which is the mass enclosed inside $b_{\rm opt}$, i.e.~the mass of the smaller part of the clump we are now considering. Since we are only considering a concentric subsphere of the original clump and this part has a smaller mass, we would also need to modify the total mass density of clumps in Eqn.~\eqref{eqn:heateqn} by replacing $\rho_{\rm clumps} \rightarrow \rho_{\rm clumps} \frac{M_{\rm enc}(b)}{M_{\rm clump}}$.  In principle we could choose any $b_{\rm opt}$ so long as it was less than $b_{\rm max}$ and it would give us some conservative lower bound on the heating rate, however we choose the optimum $b_{\rm opt}$ which maximizes the heating rate of the stars.

To summarize,  the direct heat transfer rate from clumps to stars per unit stellar mass is
\begin{align}
    H_{\star,{\rm direct}}
    =2\sqrt{2\pi} \frac{G^2 \sigma_{\rm clump}^2
    } {\left(\sigma_\star^2+\sigma_{\rm clump}^2\right)^\frac{3}{2}}\rho_{\rm clumps}(t) \frac{M_{\rm enc}(b_{\rm min})^2}{M_{\rm clump}}  \nonumber \\\times 
   \log \left(\frac{b_{90}^2+b_{\rm max}^2}{b_{90}^2+b_{\rm min}^2}
   \right) 
   \label{eqn:clumpdirectheating}
\end{align}
Here, the radial dependence of $\rho_{\rm clumps}$ is suppressed owing to the choice of a Dehnen profile with $\gamma=0$ such that the DM density is near constant inside the DM scale radius. Furthermore, as discussed in Ref.~\cite{Graham:2023unf}, the only time dependence arises from $\rho_{\rm clumps}(t)$ and to a good approximation, it can be replaced by its time average i.e.  $\rho_{\rm clumps}(t)\rightarrow \rho_{\rm clumps}(t=0)\times \eta_{\rm NT}$ which we do to simplify the computation.

The quantity $b_{\rm min}$ is given by,
\begin{align}
    b_{\rm min}=
    \begin{cases}
    b_{\rm samp} & b_{\rm samp}> r_s \\
    r_s & b_{\rm max} > r_s >b_{\rm samp}\\
    b_{\rm opt} & r_s > b_{\rm max}
\end{cases}
\label{eqn:bmin}
\end{align}
 
To get a feeling for where the optimization of $b_{\rm opt}$ pushes us in the case of $r_s > b_{\rm max}$, let us estimate $b_{\rm opt}$. The heating rate is maximized when $M_{\rm enc}^2(b)\times \log\left(\frac{b_{90}^2+b_{\rm max}^2}{b_{90}^2+b^2}\right)$ is maximized. For the density profile in Eqn.~\ref{denseqn}, we find $b_{\rm opt}\approx\frac{b_{\rm max}}{\sqrt{e}}\approx0.6b_{\rm max}$ when $b_{\rm max}\gg b_{90}$ and $b_{\rm opt}\approx 0.7 b_{\rm max}$ when $b_{\rm max}\ll b_{90}$.  So in other words, the optimum pushes us to take new, smaller clumps which are roughly the same size as the stellar scale radius.  Any smaller than this and we would gain logarithmically in the range of impact parameters allowed for the scattering but lose like a power law in the mass of the smaller clumps being considered.  So the optimum is pushed up very close to the maximum impact parameter $b_{\rm max}$.

\subsection{Tidal heating rate of stars by clumps}

In this subsection we calculate the tidal heat transfer rate from the clumps to the stars.  By tidal heating we mean scatterings where the clump passes outside of the stellar scale radius and so their dominant effect on all the stars is the same.  Thus they primarily cause a change to the center-of-mass of the whole stellar population, but not internal heating of the gas of stars.  In this subsection we calculate this suppressed heating rate.

We follow ~\cite{Graham:2023unf} and make a straight line approximation for clumps passing by and causing tidal heating of the star system. If a clump with mass $M_{\rm clump}$ passes by with impact parameter $b$ (distance from the center of the stellar distribution), the energy injected into the star cluster by one pass of a clump per unit stellar mass is~\cite{vandenBosch:2017ynq},
\begin{align}
\frac{\langle\Delta E_{\rm dt}(b)\rangle}{M_s} =\frac{4G^2 M_{\rm clump}^2}{3v_{\rm rel}^2}\langle r^2 \rangle_\star  
A_{\rm corr}(b)
\frac{\chi_{\rm st}(b)}{b^4}, 
\label{tidalheating}
\end{align}
Here $\langle r^2 \rangle_\star$ is evaluated over the star configuration (defined in Ref.~\cite{Graham:2023unf}), and $A_{\rm corr}$ is the adiabatic correction factor since the stars are in orbit. $v_{\rm rel}$ is the relative velocity between stars and clumps. The function $\chi_{\rm st}$ depends on the density profile of the clump and is described in Ref.~\cite{vandenBosch:2017ynq}. 
As a result the heating rate can be determined as,
\begin{align}
H_{\star, \rm dt}=\int_{b_{\rm min}}^{b_{\rm max}} \Delta E_{\rm dt}(b) \,  \, 2\pi b \, db \, \frac{ \rho_{\rm clumps}(t)}{M_{\rm clump}} v_{\rm clump} 
\label{eqn:hdt}
\end{align}
Similar to the direct heating case, we drop out the radial dependence as well as the time dependence of $\rho_{\rm clumps}$ via the replacement $\rho_{\rm clumps}(t)= \rho_{\rm clumps}(t=0)\times \eta_{\rm NT}$, which is a good approximation and simplifies the computation.
The lower limit of the integral is given by, 
\begin{align}
    b_{\rm min}=\textrm{Max}\left(b_{\rm samp},3 R_{0,\star}\right)
\end{align}
and $b_{\rm max}=R_{\rm UFD}$.

\subsection{Limits without incorporating clump destruction effects}

In this subsection we use the above stellar heating rates to set limits on what clumps can exist inside a UFD.  We use the UFD Segue-I with stellar scale radius of 24.7 pc as an example.  Roughly, we will call a point ruled out if the clumps cause so much heating that in order to reproduce the currently observed stellar scale radius, the stars initially needed to start in a tight cluster of size less than 2 pc.

\begin{figure}
     \centering
         \centering
         \includegraphics[width=0.48\textwidth]{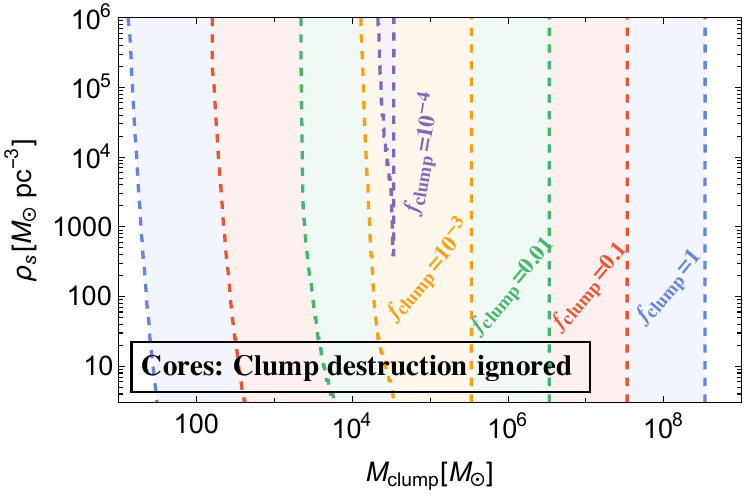}     
         \caption{This figure shows what the limits from stellar heating in the clump density vs clump mass plane would have been if there were no effects that caused destruction of clumps in UFDs.  So note that these are NOT the real limits, the real limits including destruction of clumps from tidal heating/stripping effects are shown in Figure \ref{fig:mainresults}.  We show curves for several different fractions $f_{\rm clump}$. }
         \label{fig:rhovsmbefore}
     \end{figure}

We solve the same differential equation derived in Ref.~\cite{Graham:2023unf} for the evolution of the stellar scale radius $R_{0,\star}$. 
\begin{align}
    \frac{dR_{0,\star}}{dt}=H_\star(R_{0,\star})\times\left\{\frac{3 \pi  G M_\star}{64 R_{0,\star}^2}+\frac{G M_{\rm UFD} R_{0,\star}}{4R_{\rm UFD}^3}\right.\nonumber \\\left.\left(6  \log
   \frac{R_{\rm UFD}}{R_{0,\star}}+  (3 \log (4)-17)\right)\right\}^{-1}
    \label{heatingeqndiff}
\end{align}
where $H_\star=H_{\star,{\rm direct}}+H_{\star, \rm dt}$. Setting $R_{0,\star}(\rm today)=24.73~\textrm{pc}$ and $\eta=\eta_{\rm NT}$, we set limits on the heating rate, and hence limits on the clumps by requiring that $R_{0,\star}(t=0)>2~\textrm{pc}$. A parameter point that predicts $R_{0,\star}(t=0)<2~\textrm{pc}$ is ruled out. We plot the inferred limits in the $\rho_s$ vs $M_{\rm clump}$ plane in Fig.~\ref{fig:rhovsmbefore}.  It is important to stress that these are not limits on this parameter space, since survival under clump destruction effects, covered in Sec.~\ref{clumpsurvival} are not incorporated yet. We display this figure solely to explain the dependence of the limits on the density of the clumps. 

We see that as the clump density $\rho_s$ increases towards $10^6 M_\odot~\textrm{pc}^{-3}$ we see that the boundaries become almost vertical.  On the right side the boundaries are exactly vertical and are set by sampling effects as explained in \cite{Graham:2023unf}.  On the left side the boundaries are changing only logarithmically with the density.  As the density increases at a fixed mass the clumps become smaller.  This increases the range of impact parameters allowed in the log in the heating rate, thus slightly increasing the heating rate and allowing us to rule out slightly lower masses.  But for the higher densities we consider, the low mass clumps are already quite small and this effect is small.  For lower densities we can see the lines on the left side bending due to this logarithmic effect.  If we kept plotting up to very high densities this plot would of course reproduce the results for (point-like) MACHOs/PBHs from \cite{Graham:2023unf}.
Note that fractions of DM as low as $f_{\rm clump}\sim 10^{-4}$ can be constrained.

\section{Clump survival in UFDs}
\label{clumpsurvival}

In hierarchical structure formation, the clumps formed via collapse act as particles when larger halos begin to collapse later. These clumps that are now gravitationally bound to the UFD can undergo mass loss due to various processes. We use the formalism developed in ~\cite{vandenBosch:2017ynq} to quantify clump survival due to the different processes.

In all processes of interest, we will consider a clump ``A'' transiting another object ``B''. Here ``B'' could be individual stars, other individual clumps or the host halo. 

Let the particles of clump ``A'' have individual velocities $v_i$ and $\Delta v_i$ be the velocity kick caused by the transit of ``B''. Then, the energy increase in clump ``A'' per unit mass is given by,
\begin{align}
\Delta E_i=v_i\cdot \Delta v_i +\frac{1}{2}(\Delta v_i)^2
\end{align}
The first term averages to zero, but a diffusion term survives. After $N$ encounters,
\begin{align}
\Delta E_i(N)=\sqrt{N} v_i \cdot \langle \Delta v_i \rangle + N\frac{1}{2}\langle\Delta v_i\rangle^2
\label{diffterm}
\end{align}
We will next show that it is justified to consider only the second term in order to calculate subhalo harrasment.
It was shown in ~\cite{vandenBosch:2017ynq} that $\Delta E_i \ge E_i $ is needed for the clump to get appreciable tidal heating. To be conservative we require that the total energy transferred $\Delta E_{\rm tot}\equiv \sum_i \Delta E_i$ does not exceed the binding energy $E_b\equiv|\sum_i E_i|$,  i.e. 
\begin{align}
\label{disruptionEqn}
E_b \ge \Delta E_{\rm tot}
\end{align}
in order to avoid disruption. 
Right on the edge of our disruption condition, when $\Delta E_i\sim E_i$, the two terms in equation \ref{diffterm} are comparable.
For small energy transfer $\Delta E_i$ and small $N$, the first term dominates. The second term starts becoming important when $\Delta E_i$ is comparable to the initial energy per mass, $E_i\sim v_i^2 \approx N \Delta v_i^2\sim \Delta E_i$. After this, the second term dominates. Since  Ref.~\cite{vandenBosch:2017ynq} found that $E_b \gg \Delta E_{\rm tot}$ for significant mass loss in clumps. In this limit the second term s the relevant one and hence we consider only the second term going forward.

Next let us estimate both the LHS and RHS of Eqn.~\ref{disruptionEqn}. 
The LHS is given by,
\begin{align}
E_b=4\pi G\int dr r \rho(r)M_{\rm enc}(r)
\end{align}
The RHS, for a given impact parameter $b$, is given by~\cite{vandenBosch:2017ynq},
\begin{align}
\langle\Delta E(b) \rangle=\frac{4G^2 M_B^2 M_A}{3v_{\rm rel}^2}\langle r^2 \rangle  
A_{\rm corr}
\frac{\chi_{\rm st}(b)}{b^4}
\label{ebm}
\end{align}
Here $\langle r^2\rangle$ is the density averaged $r^2$ over the clump ``A''. The function $\chi_{\rm st}$ is also defined in~\cite{vandenBosch:2017ynq} and it captures the density profile of object ``B''. It is smaller than unity for an extended mass distribution and one for a point mass. Here $A_{\rm corr}$ is the adiabatic  correction factor to account for the dilution in tidal heating when the orbital frequency of particles in ``A" about ``A"'s center is larger than the time for ``A" to transit ``B". We describe the particular form of $A_{\rm corr}$ in each individual subsection.  

In the rest of this section, we adapt this formalism for different choices of ``B''.

\subsection{Host Halo Heating}
\label{subsec:hhh}
We discuss the heating of clumps due to the gravitational potential of the host halo in this subsection. We identify ``A'' with the clumps and ``B" with the host halo. 

For a host halo that satisfies a Dehnen profile with $\gamma=0$, we can simplify the function $\chi_{\rm st}$ that captures the profile information of the host in the $b \ll R_{\rm UFD}$ limit. Taking the expression $\chi_{\rm st}$ from  \cite{vandenBosch:2017ynq}, we find in the $b \ll R_{\rm UFD}$ limit,
\begin{align}
\chi_{\rm st}(b) =\frac{b^4}{4R_{\rm UFD}^4}.
\end{align} 
Furthermore, for NFW cores, $\langle r^2 \rangle = 0.41 r_s^2$. 
Thus,
\begin{align}
\langle\Delta E_{\rm HHH} \rangle=0.41\frac{4G^2 M_{\rm UFD}^2 M_{\rm clump}}{3v_{\rm rel}^2}r_s^2
\frac{A_{\rm corr,HHH}}{4R_{\rm UFD}^4}
\label{ebm2}
\end{align}
Here the subscript HHH stands for host halo heating.

To our knowledge, thus far, the adiabatic correction for host halo heating has only been derived in the $b\approx R_{\rm host}$ limit, i.e. when the typical pericenter of the clump orbit is comparable to the scale radius of the host halo. In our case, we are interested in the survival of clumps that pass near the stars in the UFD, rather than the typical clump, since it is these clumps which are responsible for the heating of the stars.  Thus we are interested in the case $b\approx R_{0,\star} \ll R_{\rm host}=R_{\rm UFD}$ where $R_{0,\star}$ is the stellar scale radius of the UFD\footnote{Our discussion holds even in the case where pericenter changes from orbit to orbit. This is because as we show later the pericenter drops out of our calculations entirely. }.

The adiabatic correction kicks in when the timescale of scattering with the host halo $\tau$ is longer than the orbital period of particles about the clump $\frac{1}{\omega}$. $\tau$ is typically taken as $\tau=\frac{b}{v_{\rm rel}}$ where $b$ is the impact parameter. However, since the clump spends all its orbit inside the UFD, it is possible that $\tau$ is much larger, $\mathcal{O}\left(\frac{R_{\rm clumps}}{v_{\rm rel}}\right)$.  Here $R_{\rm clumps}(t)$ is the scale radius of the clump distribution which evolves in time since the clumps feel dynamical friction. To estimate this, we construct a toy model of a mass on a spring that undergoes simple harmonic motion with frequency $\omega$. This approximately captures the orbit of individual particles about the clump center. We then integrate the tidal force as the center of this harmonic oscillator traverses the UFD with pericenter $b$. For the $\gamma=0$ Dehnen profile, we confirm that the adiabatic correction is indeed very weakly dependent on $b$ but instead depends on $R_{\rm clumps}$ in the $b\ll R_{\rm clumps}$ limit. Hence we take the adiabatic factor to be
\begin{align}
A_{\rm corr,HHH}=\left(1+\frac{v_{\rm orb}^2}{r_s^2}
\frac{R_{\rm clumps}^2}{v_{\rm rel}^2}
\right)^{-\frac{3}{2}}
\end{align}
Here $v_{\rm orb}=\sqrt{\frac{0.45 G M_{\rm clump}}{r_s}}$ is the typical orbital velocity of a particle about the clump
Thus, the heating due to one passage given in Eqn.~\ref{ebm2}
is independent of the pericenter $b$.

Hence, the instantaneous heating rate for an individual clump is given by

\begin{align}
    H_{\rm HHH}(t)=\langle\Delta E_{\rm HHH}(R_{\rm clumps}(t)) \rangle \frac{v_{\rm clumps}}{R_{\rm clumps}(t)}
\end{align}

$R_{\rm clumps}(t)$ is obtained by solving the migration differential  Eqn.~\ref{heatintmacho}. 

The total amount of energy imparted after time $t$ is given by,

\begin{align}
\Delta E_{\rm HHH}^{\rm tot}(t)=\int^t_0 H_{\rm HHH}(t') dt'
\label{hhhfinal}
\end{align}

Using this quantity we can estimate the time taken for a clump to be destroyed by host halo heating, $t_{\rm HHH}$. This is the time taken for the total energy imparted to equal the binding energy, i.e. 
\begin{align}
\Delta E_{\rm HHH}^{\rm tot}(t=t_{\rm HHH})=E_b
\end{align}

We next define an effective density enhancement after incorporating host halo heating $\eta_{\rm HHH}$,

\begin{align}
\eta_{\rm HHH}=\frac{1}{t_{\rm UFD}}\int_0^{T_{\rm HHH}} dt \frac{\rho_{\rm clumps}(R_{0,\star},t)}{\rho_{\rm clumps}(R,t=0)}
\label{eqn:ehhh}
\end{align}
the difference between this equation and Eqn.~\ref{eqn:enhancement factor} is that the upper limit of the integral is now set to $t_{\rm HHH}$. 
We generally want to know the density enhancement (or reduction) in the location of the stars, so we evaluate this at the scale radius of the stars.
However once again, we can actually evaluate this anywhere inside the clumps scale radius so we replace $\rho_{\rm clumps}(R_{0,\star},t)$ with its approximate value derived in Eqn.~\eqref{eqn:coreapprox}.

$\eta_{\rm HHH}$ is plotted in Fig.~\ref{fig:enhancement} as a function of the clump mass $M_{\rm clump}$ for different choices of the density of the clump $\rho_s$ for $f_{\rm clump}=10^{-3}$ in solid lines. One can see that for core densities down to $\rho_s\approx10^3 M_{\odot}\textrm{pc}^{-3}$, $\eta_{\rm HHH}$ is identical to $\eta_{\rm NT}$ (dashed lines), i.e. the effect of host halo heating is negligible. However there is a large deviation in  $\eta_{\rm HHH}$ from $\eta_{\rm NT}$ for $\rho_s=100 \, M_{\odot}  
 \, \textrm{pc}^{-3}$ for masses $3\times 10^4 \, M_{\odot}$ and above. This can be understood as the effect of migration, which reduces $R_{\rm clumps}$ which in turn exacerbates host halo heating. Nonetheless $\eta_{\rm HHH}$ is still larger than unity. For $\rho_s=10 \, M_{\odot} \, \textrm{pc}^{-3}$, the density is so low that $t_{\rm HHH} \ll t_{\rm UFD}$ even for small clump masses where migration is negligible. As a result $\eta_{\rm HHH}$ is smaller than unity for these densities, meaning that the host halo heating effect destroys the clumps fast enough that this effect is larger than the migration effect and thus there is less total heating. 

\subsection{Clump-Clump Harassment}
The issue of subhalo-subhalo harassment was treated in detail in~\cite{vandenBosch:2017ynq}. Hence our discussion mirrors the one found in the above reference. Specifically we identify both ``A" and ``B" with clumps in this subsection. Therefore the energy imparted per clump-clump collision is given by,

\begin{align}
\langle\Delta E_{\rm CCH}(b) \rangle=0.41\frac{4G^2 M_{\rm clump}^3}{3v_{\rm rel}^2} r_s^2  
A_{\rm corr,CCH}
\frac{\chi_{\rm st}(b)}{b^4}
\label{eqn:clumpclump}
\end{align}
Here, the subscript CCH stands for clump-clump harassment. 
Here,
\begin{align}
A_{\rm corr,CCH}=\left(1+\frac{v_{\rm orb}^2}{r_s^2}
\frac{b^2}{v_{\rm rel}^2}
\right)^{-\frac{3}{2}}
\end{align}

\begin{figure}
     \centering
         \centering
         \includegraphics[width=0.48\textwidth]{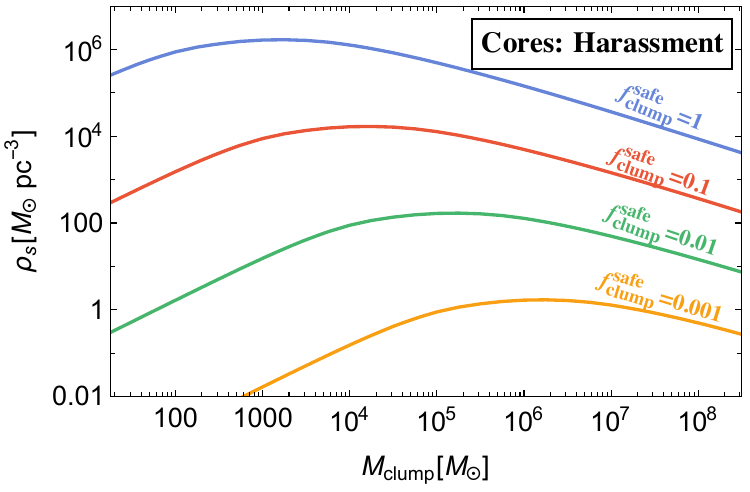}
         \caption{Contours of $f_{\rm clump}^{\rm safe}$ are plotted in the $\rho_s$ vs $M_{\rm clump}$ plane. Scale densities $\rho_s$ above a particular $f_{\rm clump}^{\rm safe}$ contour are safe from clump-clump harassment for $f_{\rm clump}^{\rm safe}=\eta_{\rm HHH}f_{\rm clump}$.}
    \label{fig:fsafe}
     \end{figure}

The total energy imparted over the age of the UFD is then,
\begin{align}
    \Delta E_{\rm tot,CCH}=\int_{b_{\rm min}}^{b_{\rm max}} &\langle\Delta E_{\rm CCH}(b) \rangle 2\pi b db \nonumber \\ &\times\eta_{\rm HHH} f_{\rm clump} \frac{\rho_{\rm DM}}{M_{\rm clump}} T_{\rm UFD}v_{\rm rel}
\end{align}
Note that we use the enhancement $\eta_{\rm HHH}$ that was evaluated at the scale radius of the stars which was shown to be a good approximation for any radius smaller than $R_{\rm clumps}$. Since a majority of the clumps sit inside $R_{\rm clumps}$ as well, $\eta_{\rm HHH}$ is an excellent approximation for the enhancement in clumps that any individual clump sees. 
Here $b_{\rm min}$ is given by solving $\pi b_{\rm min}^2 \eta_{\rm HHH} f_{\rm clump} \frac{\rho_{\rm DM}}{M_{\rm clump}} v_{\rm rel}=1$ and $b_{\rm max}=R_{\rm UFD}$. 
Unlike the other processes discussed thus far, subhalo harassment depends on the fraction of DM in clumps i.e. $f_{\rm clump}$. Afterall,  
$\Delta E_{\rm tot,CCH} =0$ for $\eta_{\rm HHH} f_{\rm clump}=0$ and monotonically increases with $\eta_{\rm HHH}f_{\rm clump}$. Hence a $\eta_{\rm HHH}f_{\rm clump}$ exists for which  $\Delta E_{\rm tot,CCH}=E_b$. We define this quantity as $f_{\rm clump}^{\rm safe}$, i.e. the maximum effective DM fraction $\eta_{\rm HHH}f_{\rm clump}$ that is safe from clump-clump harassment. 
We plot contours of $f_{\rm clump}^{\rm safe}$ in the $\rho_s$ vs $M_{\rm clump}$ plane in Fig.~\ref{fig:fsafe}. Scale densities $\rho_s$ above a particular $f_{\rm clump}^{\rm safe}$ contour are safe from harassment for $f_{\rm clump}^{\rm safe}=\eta_{\rm HHH}f_{\rm clump}$. In the small $M_{\rm clump}$ limit, the heating rate is proportional to $M_{\rm clump}$ and hence larger $\rho_s$ are required to keep the clumps harassment free for larger masses. However, at some large mass that is  $f_{\rm clump}$ dependent, the low sampling rate leads to large impact parameters which in turn leads to a large adiabatic correction. This dilutes the clumps harassment rate which leads to the turn around behavior. We see that for $f_{\rm clump}^{\rm safe}=0.01$ scale densities $\rho_s=0.01~M_{\odot}\textrm{pc}^{-3}$ are safe from harassment for the entire range of clump masses we consider. 

Clump-clump harassment is incorporated by taking the minimum of $\eta_{\rm HHH}f_{\rm clump}$ and $f_{\rm clump}^{\rm safe}$ for any parameter point $\{\rho_s,M_{\rm clump}\}$, i.e the effective clump fraction we take never exceeds the safe fraction with regard to clump-clump harassment. The final result of this section is that we have derived a replacement rule 
\begin{align}
\rho_{\rm clumps}(t)\rightarrow \rho_{\rm DM} \times \textrm{Min}\left(\eta_{\rm HHH} f_{\rm clump},f_{\rm clump}^{\rm safe}\right)
\end{align}
which should be substituted in Eqn. \eqref{eqn:clumpdirectheating} for direct heating and Eqn.\eqref{eqn:hdt} for tidal heating\footnote{Note that the enhancement factor $\eta_{\rm HHH}$ technically applies only inside the clumps' scale radius while the tidal heating comes from clumps at all distances.  However the contribution to the tidal heating coming from clumps outside the tidal radius is negligible here, partly because everywhere that tidal heating is even relevant (see Fig.~5 Right of \cite{Graham:2023unf}) the clumps' tidal radius remains always much bigger than 24.7~pc.} respectively.

We also calculated harassment from individual stars but it is far subdominant to all other clump destruction processes so we do not discuss it further.

\subsection{Effects of clump destruction on constraints}

In this subsection we show the effects that these clump destruction processes have on our constraints.  Without these processes we would have set the constraints in Figure~\ref{fig:rhovsmbefore}.  However these destruction processes cut off the curves below some density.

\begin{figure}
     \centering
         \centering
         \includegraphics[width=0.48\textwidth]{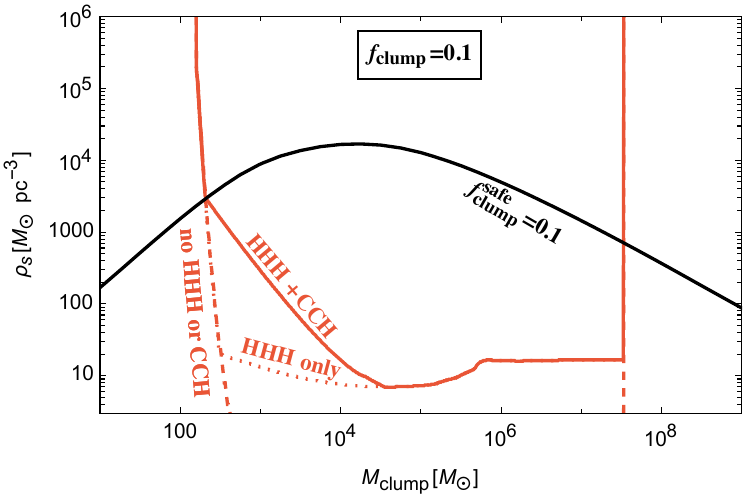}
         \caption{Limits on clump parameter space after incorporating various clump destruction processes, taking $f_{\rm clump} = 0.1$ as an example. The ``no HHH or CCH" dashed line contains no clump destruction and is identical to Fig.~\ref{fig:rhovsmbefore}.  The ``HHH only" dotted line incorporates only host halo harassment. The ``HHH+CCH" solid line additionally incorporates clump-clump harassment. The $f_{\rm clump}^{\rm safe}=0.1$ black line is identical to the red line in Fig.~\ref{fig:fsafe} and reproduced here for reference. The solid red line heres is reproduced as the solid red line in Fig.~\ref{fig:mainresults} along with contours for other $f_{\rm clump}$.}
         \label{fig:inter}
     \end{figure}

We demonstrate the effect of these clump destruction processes for a single value of $f_{\rm clump}$ in Fig.~\ref{fig:inter}.

First, the constraint that would be derived if all the clumps were deemed to survive without incorporating tidal heating processes is shown in dashed red. This is the same red dashed line from Fig.~\ref{fig:rhovsmbefore}. Next, we show the contour after incorporating host halo heating alone in dotted line (labeled HHH only). We see that this roughly cuts of the constraint at $\rho_s\approx 10 M_{\odot} \textrm{pc}^{-3}$. Finally, incorporating both host halo heating as well as clump-clump harassment (CCH) produces the solid red curve. We see that CCH is not a significant process as compared to HHH for $M_{\rm clump}\gtrsim 10^{4.5} M_{\odot}$. This is the mass range where host halo heating significantly affects $\eta_{\rm HHH}$ as seen in Fig.~\ref{fig:enhancement} owing to significant migration. Coincidentally, it is also the mass range where CCH peaks as seen in Fig.~\ref{fig:fsafe}. For smaller masses where CCH dominates, we see a further dent in constraints for ``HHH+CCH" as compared to the ``HHH only" curve. We also replot the $f_{\rm safe}^{\rm clump}=0.1$ contour from Fig.~\ref{fig:fsafe} to illustrate the fact that CCH is an issue only below the black contour. This is the reason why the ``no HHH or CCH" (dashed) curve and the ``HHH + CCH" (solid) curve share a common intersection point with the $f_{\rm safe}^{\rm clump}=0.1$ curve (black).  
The solid lines after incorporating both host halo heating as well as clump clump harassment are the main result of this paper. We discuss only these contours moving forward. Below we discuss contours for many values of $f_{\rm clump}$.  

\section{Results}
\label{results}

\begin{figure*}
     \centering
         \centering
         \includegraphics[width=0.48\textwidth]{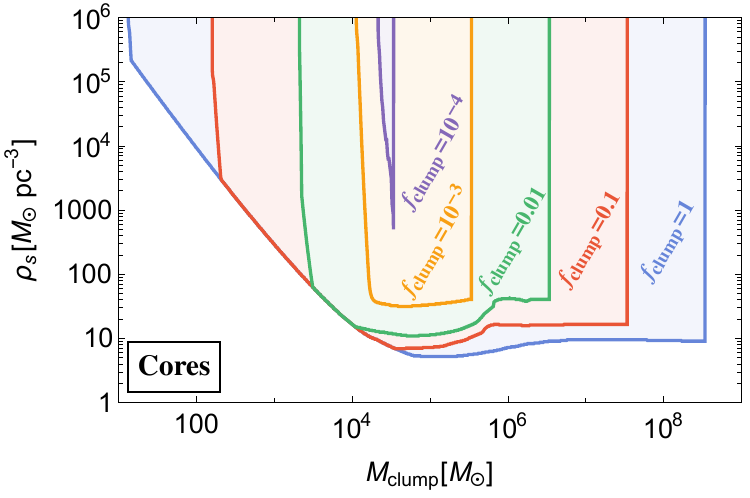}
         \includegraphics[width=0.48\textwidth]{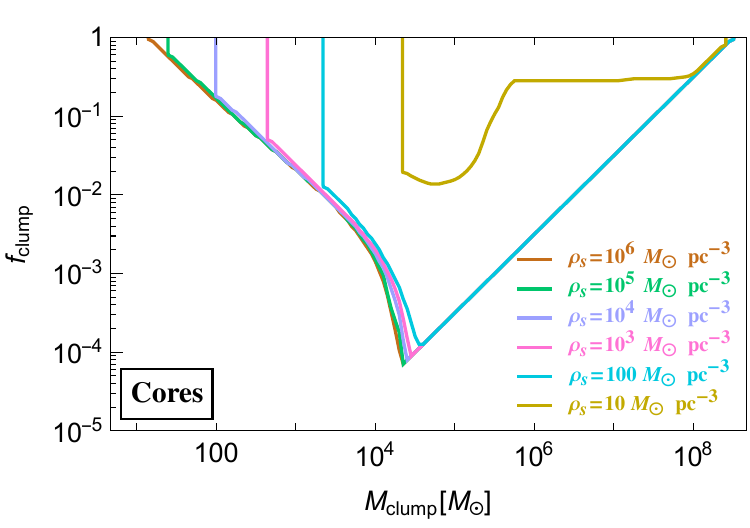}
         \caption{Our constraints on clump parameter space in clump mass $M_{\rm clump}$, clump scale density $\rho_s$, and fraction of dark matter made up of clumps $f_{\rm clump}$.  In the left plot, the different lines represent contours of different fraction $f_{\rm clump}$ in the $\rho_s$ - $M_{\rm clump}$ plane.  In the right plot the different lines represent contours of different $\rho_s$ in the $f_{\rm clump}$ - $M_{\rm clump}$ plane.}
         \label{fig:mainresults}
     \end{figure*}

In this section we present our results for constraints on the DM subhalos/clumps in a UFD arising from their heating effect on the stellar population.  We use the heating rate calculation given in Section \ref{clumpheat} to find the parts of clump parameter space (clump mass, density, and dark matter fraction) which produce excessive heating, assuming such clumps exist in the UFD.  This produces the region shown in Figure~\ref{fig:rhovsmbefore} where, if such clumps existed in the UFD, they would have caused too much heating of the stars.  We then combine this with the calculations of Section~\ref{clumpsurvival} which determine the minimum density necessary for a clump to survive destruction in the environment of the UFD (as a function of clump mass and fraction of dark matter).  This puts a lower bound on the regions of Figure~\ref{fig:rhovsmbefore} where clumps could exist and thus cuts off our constraints at the lowest densities. 
Combining these gives our main result shown in Figure~\ref{fig:mainresults}.

\begin{figure*}
     \centering
         \centering
         \includegraphics[width=0.48\textwidth]{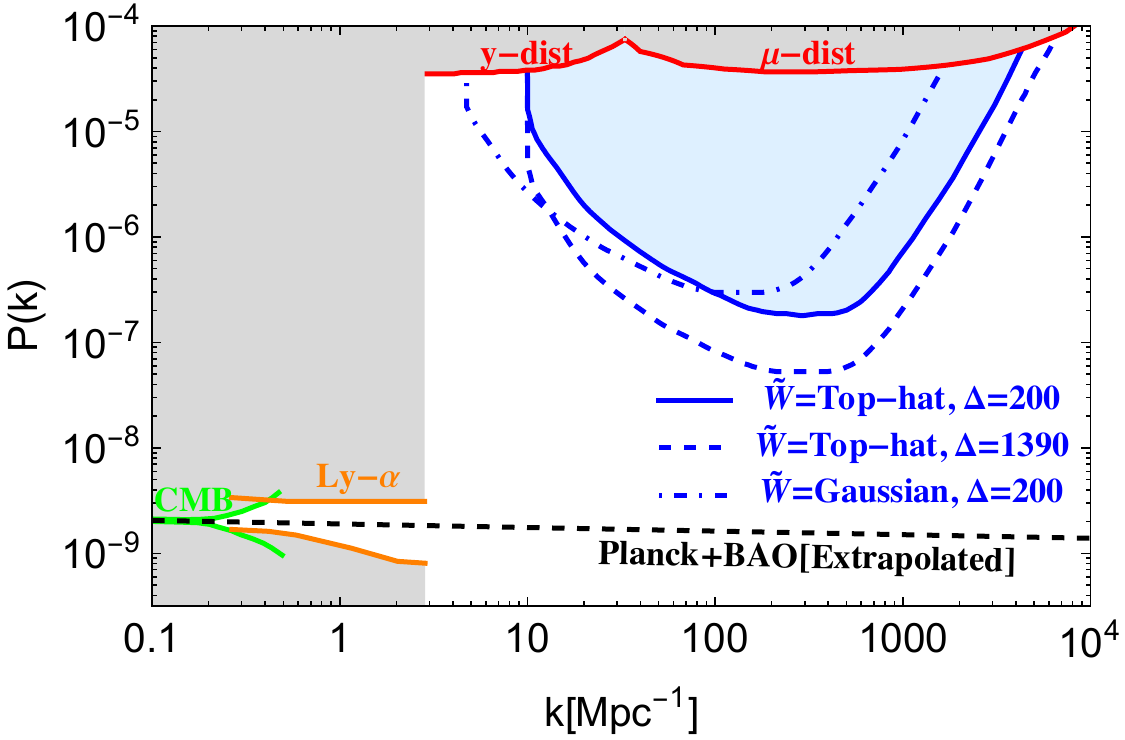}
          \includegraphics[width=0.48\textwidth]{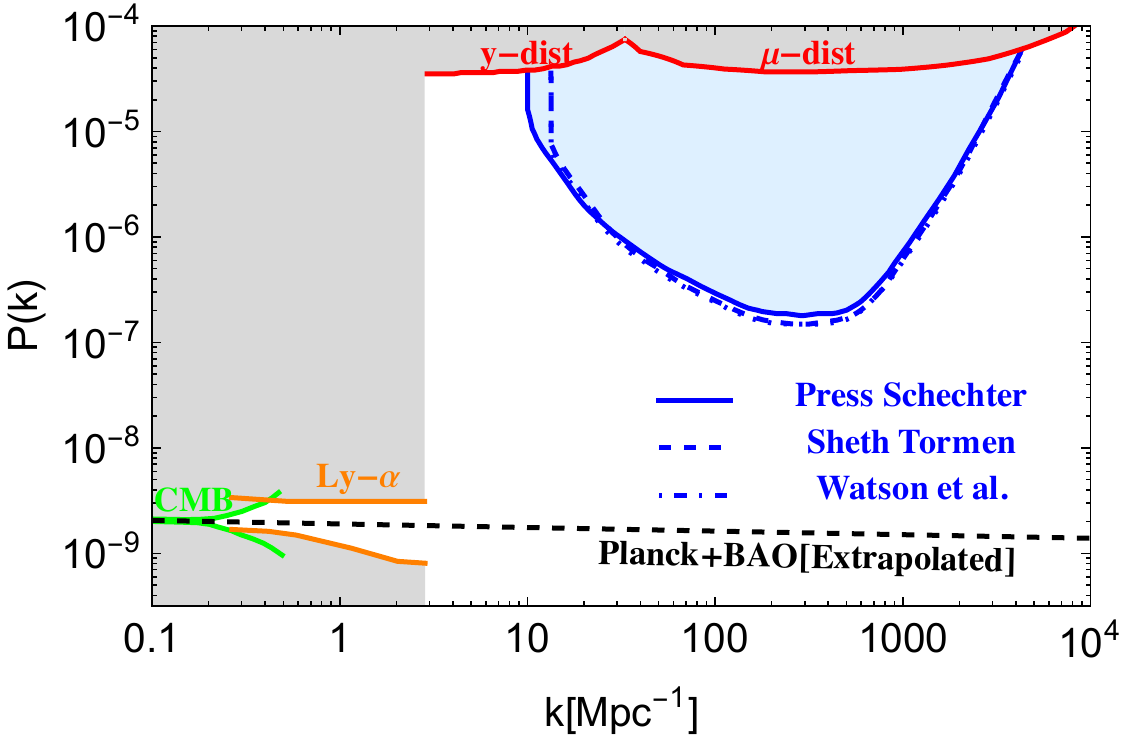}
         \caption{ Our limits (blue) on the  power spectrum of adiabatic primordial density perturbations $P(k)$ vs the wave-number $k$. These are significantly stronger than existing bounds in this range from FIRAS (red) via $\mu$-distortions and y-distortions~\cite{Chluba:2012we}.  At longer length scales the power is measured by CMB (green) \cite{akrami2020planck} and Lyman-$\alpha$ (orange) \cite{bird2011minimally}. For comparison we also show the nearly scale invariant power measured by Planck extrapolated to higher $k$ (black-dashed). The blue shaded region corresponds to the top-hat window function with the Press-Schechter~\cite{press1974formation} multiplicity function and $\Delta=200$. In the \textbf{Left} panel, we illustrate the uncertainty associated with a choice of filter, the Gaussian (dot-dashed blue), as well as an alternate choice of $\Delta=600$. In the \textbf{Right} panel, we show alternate choices of the multiplicity function: Sheth-Tormen~\cite{sheth1999large} (Dashed) and Watson et al.~\cite{watson2013halo} (Dot-dashed).}
         \label{fig:powerspectrum}
     \end{figure*}

\subsection{Limits on clumps in UFDs}

Figure~\ref{fig:mainresults} shows our constraints on the clump parameter space in the three variables: clump mass $M_{\rm clump}$, clump density $\rho_s$, and fraction of dark matter made up of these clumps $f_{\rm clump}$.

The left plot of Figure~\ref{fig:mainresults} appears similar to Figure~\ref{fig:rhovsmbefore} except the limits are cut off below a certain density because clumps of such low densities are destroyed by tidal forces in the UFD before they can heat the stars excessively. 
For masses above $\sim 10^5 \, M_\odot$ the lower limit on the density that we can constrain comes from host halo heating.  For masses below $\sim 10^5 \, M_\odot$ the lower limit arises from clump-clump harassment.
The constraints cut off sharply at high masses (the vertical edges on the right side of the curves) because of sampling; past those edges there simply are not any clumps of that size in the UFD.  The left edge of each curve at low masses comes from the heating rate calculation, at masses below those edges such clumps would not heat the stars significantly.

The right plot of Figure~\ref{fig:mainresults} is similar to the main results in Figure 6 of Paper 1~\cite{Graham:2023unf}.  Thus we see that for clumps of high enough density (above $\sim 10^6 \, M_\odot \, \text{pc}^{-3}$) there is essentially no difference from the point mass (MACHO/PBH) case.  Such dense clumps do not get destroyed by the environment of the UFD.  And the fact that they are spatially extended (much more so than a PBH for example) makes almost no difference to the heating rate since gravitational scattering is inherently long-range and the only loss is a small reduction in the log (see Eqns.~\eqref{eqn:clumpdirectheating} and \eqref{eqn:bmin} for example). As $\rho_s$ is dialed down, we see there is an abrupt drop in limit at a small enough $M_{\rm clump}$ . This is due to clump-clump harassment and occurs when the contour hits $f_{\rm clump}=f_{\rm safe}$. For much smaller densities $\rho_s\approx 10~M_{\odot}~\textrm{pc}^{-3}$, we finally see the effect of host halo heating which further suppresses the constraints.

\subsection{Primordial Power Spectrum}

We can now translate our limits on the existence of clumps in UFDs into limits on the primordial power spectrum of density perturbations from inflation.
The primordial power spectrum determines the distribution of masses and densities of dark matter structure that forms in the early universe.
There are of course significant uncertainties in predicting the spectrum of clumps in a UFD that will arise from a given primordial power spectrum of density perturbations.
For simplicity we will take an analytic procedure for estimating this, and try to keep our estimated limits on the primordial power spectrum as conservative as possible.

Current observations are consistent with a nearly scale invariant primordial power spectrum upto comoving wave-number $k\approx3\,\textrm{Mpc}^{-1}$
\begin{align}
    P_{\rm CMB}(k)= P(k_*) \left(\frac{k}{k_*}\right)^{n_s-1}
    \label{eqn:planckbao}
\end{align}
with $\log \left(10^{10} P(k_*)\right)=3.047\pm 0.014$ and $n_s=0.966\pm 0.004$ with $k_*=0.05~\textrm{Mpc}^{-1}$ obtained from Ref.~\cite{Planck:2018vyg}. Limits on low mass halos from strong lensing was used to constrain power down to the $10~\textrm{Mpc}^{-1}$ scale in  Ref.~\cite{gilman2022primordial}.  There are only very weak limits $P_k\approx10^{-4}$ at larger scales arising from $y$ and $\mu$ distortions of the CMB as measured by FIRAS~\cite{Chluba:2012we}. 

Here we translate the derived limits on clumps onto limits on the primordial power spectrum at higher scales. We choose to place constraints on a power spectrum which is the nearly flat spectrum given above plus a spike in the spectrum at some $k$ which is higher than the range of observations.

So we model the excess power on small scales as a delta function at scale $k_{\rm high}$
\begin{align}
    P(k)=P_{\rm CMB}(k)+P_k k_{\rm high} \delta(k-k_{\rm high})
    \label{eqn:powerspectrumform}
\end{align}
We will then place limits on the amount of power at short scales in the $P_k$ vs $k_{\rm high}$ parameter space.

Given a primordial power spectrum $P(k)$, we can obtain the linear matter spectrum $P_{\rm lin}(k)$ via,
\begin{align}
P_{\rm lin}(k)=\frac{1}{4}2\pi^2 P(k) T(k)^2 k\left(\frac{1}{k_{\rm eq}}\right)^4
\end{align}
Here, $k_{\rm eq}$ is the wave-number of the perturbation mode that enters the horizon at matter-radiation equality and $T(k)$ is the transfer function, which we take from Eisenstein and Hu~\cite{eisenstein1999power}. We can next calculate the variance, 
\begin{align}
\sigma^2(R,z)=\frac{D^2(z)}{2\pi^2}\int k^2P_{\rm lin}(k) \tilde{W}^2(k R) dk   
\end{align}
where $D(z)$, is the growth function, which we obtain from Ref.~\cite{diemer2018colossus} and $\tilde{W}(kR)$ is the filter function and $R$ is the comoving length scale. 

The filter functions we consider are:
\begin{align}
  \tilde{W}_{\rm tophat}(x)&=\frac{3}{x^3}\left(\sin(x)-x \cos(x)\right)\nonumber \\
  \tilde{W}_{\rm gaussian}(x)&=e^{-x^2/2}
\end{align}

The halo mass function at a redshift $z$ is given by,
\begin{align}
    \frac{dn_{\rm clump}(z)}{d\ln M_{\rm clump}}=f(\sigma) \frac{\rho_M(z)}{M_{\rm clump}}\frac{d\ln \sigma^{-1}}{d\ln M_{\rm clump}}
\end{align}
Here, $n_{\rm clump}$ is the number density of clumps, with clump mass $M_{\rm clump}$ and $\rho_M(z)$ is the non-relativistic matter density at redshift $z$.  Since $R$ is a comoving scale, we set $M_{\rm clump}=\frac{4}{3}\pi R^3 \rho_{\rm crit}\Omega_m$, where $\rho_{\rm crit}$ is the energy density of the universe today and $\Omega_m$ is the matter fraction today. This can be used to write $\sigma$ as a function of $M_{\rm clump}$. The multiplicity function $f(\sigma)$ is chosen to be that given by Press-Schechter~\cite{press1974formation} for limit setting purposes. We also provide limits with other choices such as Sheth-Tormen~\cite{sheth1999large} and Watson et al~\cite{watson2013halo} in order to show how much our limits depend on this choice. 

Now that we have the number density given a primordial power spectrum, we can compare it to the limits derived in the previous section, to finally obtain the limits on primordial power. 
Our procedure for setting this limit is as follows.
Given a primordial power spectrum $P(k)$, we obtain the fraction of DM that has collapsed into clumps with masses within a decade of $M_{\rm clump}$ mass  by redshift $z$ as 
\begin{align}
f_{\textrm{clump},P(k)}(M_{\rm clump},z)\approx M_{\rm clump}\frac{dn_{\rm clump}(z)}{d\ln M_{\rm clump}}
\end{align}

For a clump that collapsed at redshift $z$, its average density is given by
\begin{align}
\langle \rho_{\rm clump} \rangle(z) \approx \Delta \rho_{\rm crit}\Omega_m (1+z)^3 
\end{align}
Here $\Delta \approx 200$ is the resultant over density above the ambient matter density due to gravitational collapse. If we assume that this results in NFW cores, then the density at the scale radius of the clump is
\begin{align}
\rho_s(z) \approx \frac{\Delta}{2.317} \rho_{\rm crit}\Omega_m (1+z)^3 
\end{align}

The limits we set in Fig.~\ref{fig:mainresults} give a maximal allowed fraction $f$ as a function of mass and density, namely $f_{\rm clump,lim}\left(M_{\rm clump},\rho_s\right)$.
Thus, we consider a $P(k)$ distribution ruled out if there exists any $z$ and any $M_{\rm clump}$ for which,
\begin{align}
f_{\textrm{clump},P(k)}\left(M_{\rm clump},z\right) > f_{\rm clump,lim}\left(M_{\rm clump},\rho_s(z)\right)
\end{align}
So to decide whether a particular $P(k)$ is ruled out, we scan over all $z$\footnote{Actually we only include redshifts that are before the time of formation of a UFD which we assume to be around $z \sim 6 - 10$.  However this constraint does not affect our result at all because we will only use much higher redshifts in order to get to the clump densities that our results can constrain.} and all $M_{\rm clump}$.

If a particular $P(k)$ from Eqn.~\eqref{eqn:powerspectrumform} is ruled out, then we will say that value of $P_k$ is ruled out at the scale $k=k_{\rm high}$.

We plot the limits on $P_k$ vs $k$ obtained this way as the blue curve in Fig.~\ref{fig:powerspectrum}.

Our limits on the  power spectrum of primordial density perturbations $P_k$ vs the wave-number $k$ are more than two orders of magnitude stronger than existing bounds in this range from FIRAS (red) via $\mu$-distortions and y-distortions~\cite{Chluba:2012we}. They extend beyond wavenumbers $k\approx \textrm{kpc}^{-1}$.  At longer length scales, the power is measured by CMB (green) \cite{akrami2020planck} and Ly$\alpha$ (orange) \cite{bird2011minimally}. For comparison we also show $P_{\rm CMB}$ from Eqn.~\ref{eqn:planckbao} i.e.~the nearly scale-invariant Planck constrained power extrapolated to much higher $k$ in black-dashed.  The blue shaded region corresponds to the top-hat window function with the Press-Schechter multiplicity function and $\Delta=200$. 

In the left panel, we illustrate the uncertainty associated with the choice of filter.  A Gaussian filter (dot-dashed blue)  shifts the limits somewhat to lower $k$. While we have chosen $\Delta=200$ to be conservative, far higher values of $\Delta$ are found in literature~\cite{VanTilburg:2018ykj,Gorghetto:2022sue}. To illustrate the constraining power for a larger $\Delta$, we show the limit for a top-hat window and $\Delta=1390$ (which corresponds to $C_\rho=600$ in Ref.~\cite{VanTilburg:2018ykj}) in dashed lines. As expected, this puts deeper constraints on the primordial power. 

In the right panel, we show alternate choices of the multiplicity function: Sheth-Tormen (Dashed) and Watson13 (Dot-dashed). As seen, the constraints are fairly independent of the choice of multiplicity function.

\subsection{Isocurvature Perturbations}

\begin{figure}
     \centering
         \centering
         \includegraphics[width=0.48\textwidth]{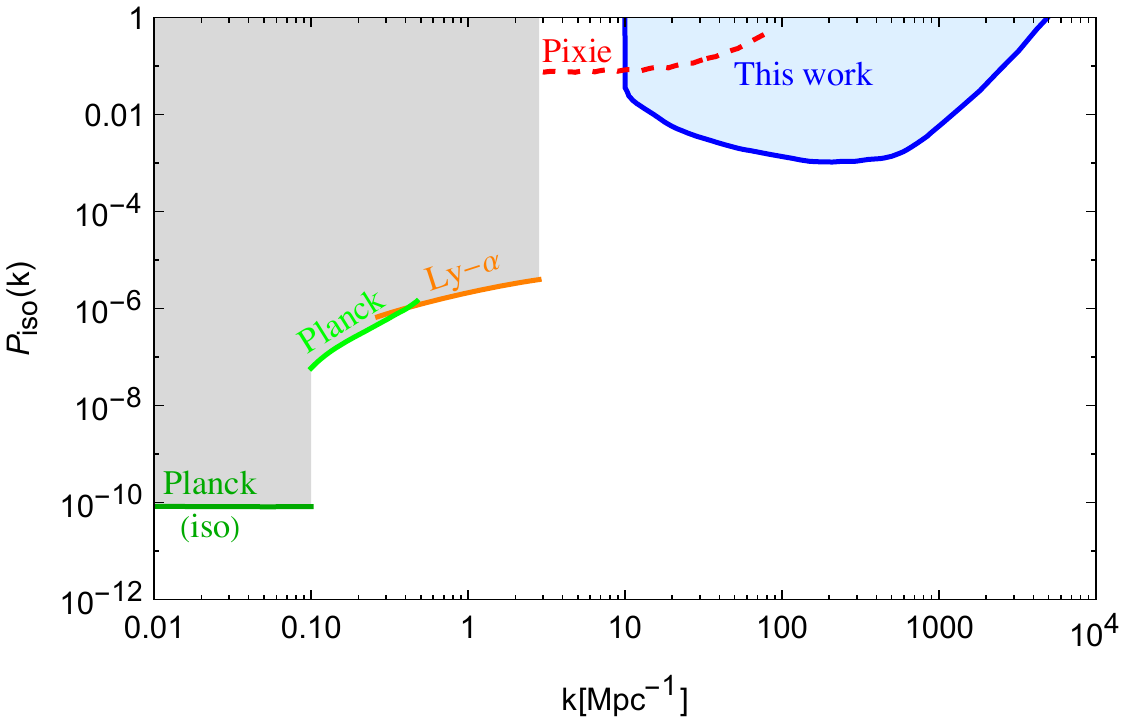}
         \caption{Our limits (blue) on the  power spectrum of dark matter isocurvature primordial density perturbations $P_k$ vs the wave-number $k$. These are the only bounds we know in this range of scales. Also shown are the PIXIE projections via $\mu$-distortions obtained by appropriately rescaling the projections for the adiabatic power spectrum found in ~\cite{Chluba:2012we} (Refer to text for more details).  At longer length scales the isocurvature power cannot exceed the power measured in the CMB by Planck (green) \cite{akrami2020planck} and Lyman-$\alpha$ (orange) \cite{bird2011minimally}. At even longer scales, the limits on isocurvature set by Planck Ref.~\cite{akrami2020planck} are shown (dark green).}
    \label{fig:piso1}
     \end{figure}

Our limits apply not just to adiabatic density perturbations as discussed above, but also to isocurvature perturbations in the dark matter density alone.  Anything that causes dark matter to form structures at the relevant scales $\sim 10 - 10^8 \, M_\odot$ will be constrained by our approach.  We show our limit on the power spectrum of dark matter isocurvature perturbations $P_{\rm DM, iso}$ in Figure~\ref{fig:piso1}.  Again we have taken a delta function spike at $k$ for the DM isocurvature power spectrum to make this plot.

The solid blue line in Figure~\ref{fig:piso1} for limits on the isocurvature power spectrum is higher than the solid blue line in Fig.~\ref{fig:powerspectrum} for the adiabatic power spectrum.  This is because isocurvature perturbations do not have log growth prior to matter-radiation equality, unlike adiabatic perturbations which do grow logarithmically.  Thus the transfer function for isocurvature perturbations lacks the log growth factor that is in the adiabatic transfer function (see e.g.~\cite{Blinov:2021axd}).

We also show existing constraints on the dark matter isocurvature power spectrum arising from Planck \cite{akrami2020planck} (dark green), which require dark matter isocurvature to be no more than $4\%$ of the power in the adiabatic density perturbations at those scales.  Further, as a rough approximation we assume the dark matter isocurvature perturbations cannot be larger than the measurements of the total power spectrum in adiabatic perturbations from CMB (light green)~\cite{akrami2020planck} and Ly$\alpha$ (orange) \cite{bird2011minimally}, however these two curves are also modified from their values in Figure~\ref{fig:powerspectrum} because of the different transfer function for isocurvature perturbations.

Note that the FIRAS constraints on the adiabatic perturbations do not appear in Figure \ref{fig:piso1}.  The FIRAS constraints are actually constraints on the photon power spectrum $P_\gamma$.  Dark matter isocurvature perturbations will alter the photon power spectrum at horizon reentry on the corresponding scales, but this is suppressed by $P_\gamma (k \approx a H) \sim \mathcal{O}(10) \frac{k_{\rm eq}^2}{k^2} P_{\rm DM, iso}$ \cite{Redi:2022llj} where $k_{\rm eq}\sim0.01\,{\rm Mpc}^{-1}$ is the comoving scale of matter-radiation equality.  Thus the FIRAS constraint is suppressed significantly and does not appear on the plot.  A projection for the sensitivity of PIXIE to dark matter isocurvature perturbations \cite{Chluba:2013dna} is shown in Figure~\ref{fig:piso1} which we derived using the same suppression factor as for FIRAS.  

We are not aware of any other constraints besides ours on dark matter isocurvature perturbations in the same range of scales as our limits.

\begin{figure}
     \centering
         \centering
         \includegraphics[width=0.49\textwidth]{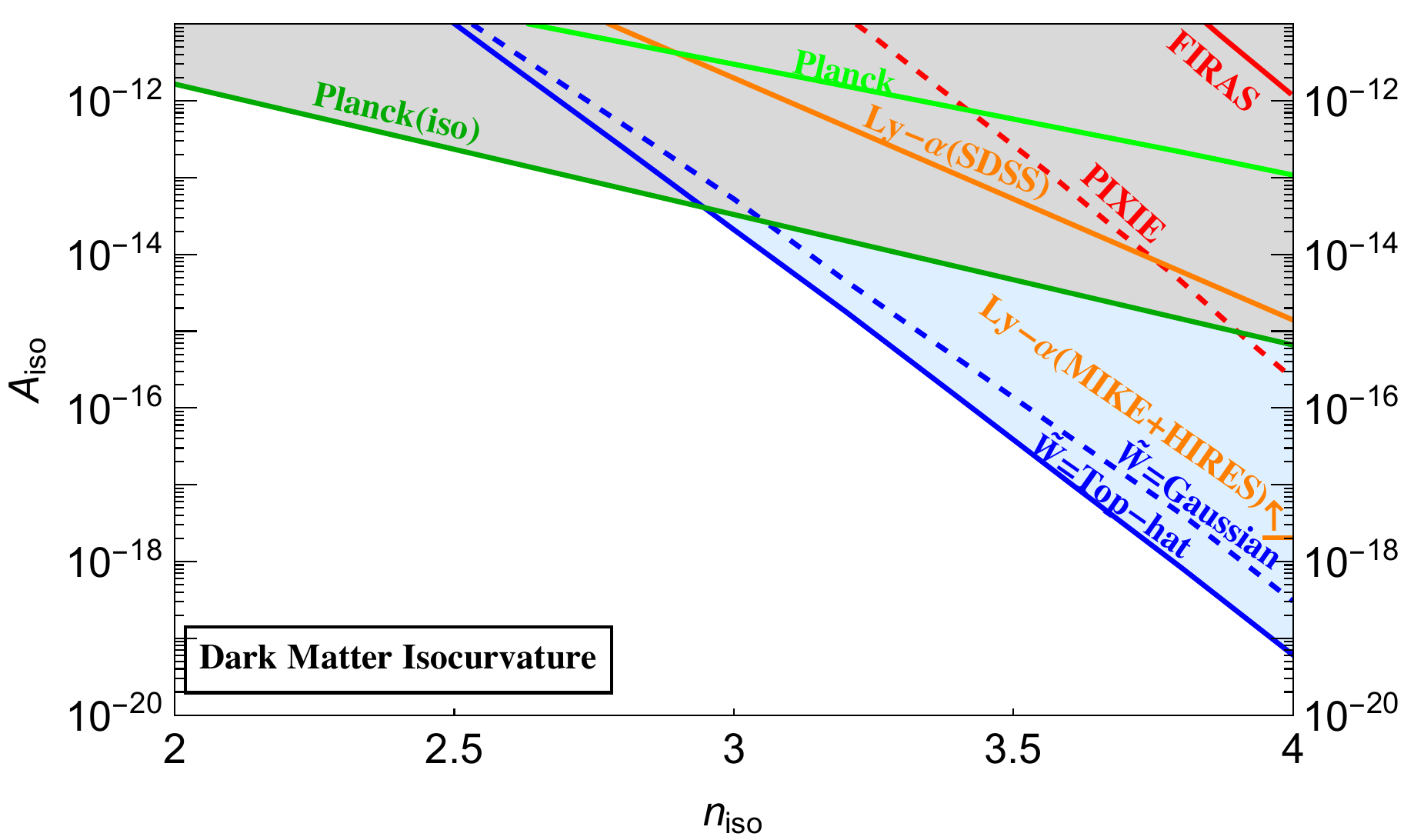}
         \caption{Our limit (blue) on a power law  isocurvature dark matter power spectrum $P_{\rm DM, iso} = A_{\rm iso} \left( \frac{k}{k_0} \right)^{n_{\rm iso} -1}$ is shown in the $A_{\rm iso}$ vs $n_{\rm iso}$ plane. We calculate existing limits from Planck's isocurvature measurement (dark green) and adiabatic power constraints (light green) \cite{akrami2020planck} and Lyman-$\alpha$ (orange), see text for more details.  Limits from FIRAS (red) \cite{Chluba:2013dna} as well as projections from PIXIE (red dashed) \cite{Chluba:2013dna} are also shown.}
    \label{fig:piso2}
     \end{figure}

We can also ask how our results constrain a power law spectrum for the DM isocurvature perturbations rather than a delta function spike at a given $k$.  We take the parametrization of \cite{Chluba:2013dna}
\begin{align}
P_{\rm DM, iso} = A_{\rm iso} \left( \frac{k}{k_0} \right)^{n_{\rm iso} -1}
\end{align}
where $k_0 = 0.002 \, {\rm Mpc}^{-1}$.  We show limits on $A_{\rm iso}$ as a function of $n_{\rm iso}$ in Figure \ref{fig:piso2}.
Our limits are stronger than existing constraints, which arise dominantly from Planck (green) \cite{akrami2020planck} and Ly$\alpha$  (orange) observations at lower $k$.  
The Ly$\alpha$ constraint from MIKE/HIRES \cite{Murgia:2019duy,Irsic:2019iff} was only placed on a spectrum with $n_{\rm iso} = 4$ so we only show it there.
The Ly$\alpha$ constraint from SDSS \cite{bird2011minimally} is obtained by requiring that $P_{\rm DM,iso}(k)$ is smaller than the SDSS Ly$\alpha$ limit at $k\approx3~\textrm{Mpc}^{-1}$ (as shown in Figure~\ref{fig:piso1}).
A similar procedure but invoking the isocurvature limits from Planck~\cite{akrami2020planck} at $k=0.1~\textrm{Mpc}^{-1}$ produces the dark green line.  And finally a similar procedure using Planck's constraints on adiabatic power~\cite{akrami2020planck} produces the light green line. Also shown are limits from FIRAS and projections from PIXIE as derived in Ref.~\cite{Chluba:2013dna}. 

Note that when $n_{\rm iso} = 4$ the power spectrum is $\propto k^3$ and we set a limit $A_{\rm iso} < 6\times 10^{-20}$.  This is a universal, white noise form that arises in many production mechanisms for ultralight dark matter (see e.g.~\cite{Graham:2015rva, Redi:2022llj, Amin:2022nlh}).  Our results give the strongest constraints on this generic form of the DM isocurvature spectrum which applies to many models.

\section{Discussion}
\label{sec:discussion}

We have placed limits on the existence of dark matter subhalos by the heating effect they would have on the stars in an ultrafaint dwarf galaxy, see Figure~\ref{fig:mainresults}.
This bounds subhalos of mass between about $10 \, M_\odot$ and $10^8 \, M_\odot$.  There are many ramifications of this bound.  For example we demonstrated that this places limits on the primordial power spectrum over a range of higher $k$ (shorter distances) than CMB or Ly$\alpha$ measurements can reach, see Figure~\ref{fig:powerspectrum}.  These limits are orders of magnitude stronger than existing limits in this range.  And at $k \sim 100 \, {\rm Mpc}^{-1}$ the limit we place currently with UFDs is almost as strong as the projected limit that the PIXIE mission would be able to place (see e.g.~\cite{Byrnes:2018txb, Chluba:2012we}).  Further, we also constrain dark matter isocurvature perturbations, giving the strongest limit in this range of $k$, see Figure~\ref{fig:piso1}.  Many models produce a rising power-law spectrum of dark matter isocurvature perturbations.  For such models our constraint is stronger than the Planck or Ly$\alpha$ constraints on isocurvature, see Figure~\ref{fig:piso2}.

Although these limits are not currently able to constrain a straight-line extrapolation from lower-k CMB measurements, it may be possible to improve the reach of this technique in the future.  The lowest power that we can constrain (Figure~\ref{fig:powerspectrum}) is being limited by the lowest density clumps that we can constrain (in Figure~\ref{fig:mainresults} Left).  This lower edge of the density constraints is primarily limited by what density clumps actually survive in the environment of the UFD.  In order to estimate this we have made multiple conservative assumptions.  Thus it is possible that a simulation of subhalos inside a UFD could improve these bounds.  Further, considering host objects other than UFDs might improve these bounds since the high densities of the UFD can lead to greater destruction of clumps.

Additionally, as discussed in~\cite{Graham:2023unf}, it might even be possible to look for positive signals of this heating effect from clumps on stars, thus possibly providing a way to actually detect the presence of dark matter subhalos at masses below $\sim 10^8 \, M_\odot$.  This would require a large sample of UFDs and more accurate predictions for the effect of such heating on the stellar distributions as a function of the various parameters of the UFD (density profiles, size, age, etc.).

Even without reaching the straight-line extrapolation, this significant improvement in the limits on the primordial power spectrum (or really on the existance of subhalos) can have important consequences.

For example, it was recently pointed out that our current knowledge of the power spectrum puts a lower limit on the possible mass of the dark matter particle, under certain assumptions about the DM formation mechanism \cite{Amin:2022nlh}.  Our bounds are at significantly higher $k$ and thus likely significantly strengthen this bound on the mass of dark matter.

And of course, our new bounds limit certain models of inflation which would generate additional power at these higher $k$ (there are many papers on inflationary models which generate higher power at high $k$ but for some examples see~\cite{Leach:2001zf, Pajer:2013fsa, Hertzberg:2017dkh, Mishra:2019pzq, Inomata:2021tpx, Hooper:2023nnl}).

Further, even if the initial power spectrum from inflation is flat, non-trivial dynamics of dark matter could generate more substructure than expected for standard CDM (see e.g.~\cite{Kolb:1993zz, Visinelli:2017ooc, Eggemeier:2019jsu, Arvanitaki:2019rax, Graham:2015rva, Gorghetto:2022sue, Gilman:2021sdr, Slone:2021nqd, Zeng:2023fnj, Gad-Nasr:2023gvf, Mace:2024uze, Gemmell:2023trd, Roy:2023zar, Chang:2018bgx, Curtin:2019ngc, Dvali:2019ewm}). For example axion dark matter can form  large, dense clumps \cite{Arvanitaki:2019rax}, so some axion DM parameter space can be explored/constrained using our results.
Or for example self-interacting dark matter (SIDM) is known to change dark matter structures on short scales.  SIDM can greatly increase the density of the cores of dark matter subhalos (see e.g.~\cite{Gilman:2021sdr, Zeng:2023fnj, Gad-Nasr:2023gvf, Slone:2021nqd, Mace:2024uze}).  These cores would then be much more likely to survive the various tidal destruction forces at work in the UFD environment.  Thus, our limits could put strong constraints on SIDM.  We leave consideration of this for future work.

In \cite{Graham:2023unf} we applied this heating effect to set limits on effectively point-like objects (MACHOs/PBHs).  In this paper we applied the effect to find limits on gravitationally-bound clumps of any kind of dark matter.  However, our results will also apply to extended dark objects bound together by forces stronger than gravity such as axion halos (see e.g.~\cite{Arvanitaki:2019rax, Croon:2024rmw, Wadekar:2022ymq}).  And in fact our results will be stronger when applied to such objects, assuming the binding force is stronger than gravity, since the main effect limiting our results was the tidal destruction of the dark matter clumps.  The extended size of the object does not decrease the heating effect significantly since it only affects the log as discussed above.
For objects that do not get tidally disrupted, our limits would become as strong as the curves in Figure~\ref{fig:rhovsmbefore} rather than the curves in Figure~\ref{fig:mainresults}~(left).
Our bounds are complementary to existing bounds (see e.g.~\cite{Croon:2024rmw}).
Previously there was a gap in the limits on extended dark objects  making up all of dark matter for masses above about $10\, M_\odot$ which our bounds close \cite{Croon:2024rmw}.

Studying structures smaller than currently observable dwarf galaxies is an exciting way to learn more about cosmology and dark matter.  We believe that studying the heating effect such structures have on stars is one very promising technique for learning more about such small-scale structure in our universe.

\acknowledgments

We thank M.~Amin, H.~Bagherian, Z.~Bogorad, A.~Ireland, D.E.~Kaplan, M.~Kaplinghat, P.~ Mansfield, G.~Marques-Tavares, O.~Slone, J.~Thompson, R.~Wechsler for useful discussions.

The authors acknowledge support by NSF Grants PHY-2310429 and PHY-2014215, Simons Investigator Award No.~824870, DOE HEP QuantISED award \#100495, the Gordon and Betty Moore Foundation Grant GBMF7946, and the U.S. Department of Energy (DOE), Office of Science, National Quantum Information Science Research Centers, Superconducting Quantum Materials and Systems Center (SQMS) under contract No.~DEAC02-07CH11359.

\bibliography{biblio.bib}
\end{document}